\documentclass[12pt,reqno]{article}
\usepackage{graphicx}	% Including figure files
\usepackage{amsmath}	% Advanced maths commands
\usepackage{amssymb}	% Extr

\usepackage[totalwidth=460truept,totalheight=600truept]{geometry}

\usepackage[T1]{fontenc}
\usepackage{color}

\global\arraycolsep=1truept
\newcommand{\beq}{\begin{equation}}
\newcommand{\eeq}{\end{equation}}
\newcommand{\bea}{\begin{eqnarray}}
\newcommand{\eea}{\end{eqnarray}}

\definecolor{darkred}{rgb}{.8,0,0}

\definecolor{darkblu}{rgb}{0,0,.8}

\usepackage{ulem}                 

\def\d{\displaystyle}

\begin{document}

\vskip 1.4truecm

\begin{center}

%----------------------------------------------------------------------%
%  title page
%----------------------------------------------------------------------%
\pagestyle{empty}
%\vspace*{1.0in}
%\rightline{10-28}
\vspace{1.2cm}
\begin{center}
\LARGE{\bf On cosmological expansion and local physics}
\\[12mm] 
%}\\

\vspace{9mm}

{\large
\textsc{J.~M.~Pons$^{1,2}$ and  P.~Talavera$^{2,3}$}}

\vspace{8mm}

\footnotesize{
$^1$ Department of Physics, 
University of Barcelona,\\  Carrer de Mart\'in Franqu\`es, 1, 11,
Barcelona,  08028, E\\
\medskip
$^2$Institut de Ciencies del Cosmos,
Universitat de Barcelona\\
Carrer de Mart\'in Franqu\`es, 1, 11,
Barcelona,  08028, E\\
\medskip
$^3$Department of Physics,
Polytechnic University of Catalonia,\\ Diagonal 647,
Barcelona, 08028, E
}

\vspace{4mm}
{\footnotesize\upshape\ttfamily  pons@icc.ub.edu,  pere.talavera@icc.ub.edu} \\

%\begin{abstract}
\vspace{2mm}
\noindent

\small{\bf Abstract} \\%[5mm]
\end{center}
\begin{center}
\begin{minipage}[h]{\textwidth}

We find an exact convergence in the local dynamics described by two 
supposedly antagonistic approaches in modern cosmology: one 
starting from an expanding universe perspective such as FLRW, the other based on a local model ignoring 
any notion of expansion, such as Schwarzschild dS. 
Both models are in complete agreement when the local effects 
of the expansion are circumscribed to the presence of the cosmological constant. 
We elaborate in the relevant role of static backgrounds like the Schwarzschild-dS metric
in standard form as the most proper 
coordinatizations to describe physics at the local scale. Finally, 
making use of an old and too often forgotten relativistic kinematical invariant, 
we clarify some widespread misunderstandings on space expansion,
cosmological and gravitational redshifts. As a byproduct we propose a {\sl unique and
unambiguous} prescription to match the local and cosmological expression of a specific observable.

\end{minipage}
\end{center}
\newpage
%----------------------------------------------------------------------%
%  Resetting of counters
%----------------------------------------------------------------------%
\setcounter{page}{1}
\pagestyle{plain}

\end{center}

\section{Motivation}

Different and opposing views coexist at this moment as to whether the expansion of the universe affects
the local dynamics at the scale of the solar system, and the amount of this effect. 
It is true that the consequence, if any, is too small 
to be detectable, but the question of principle remains as to what impact has the expansion of the 
universe on local systems. Here we focus on two --in principle-- opposite views that
compete in this arena:

{\sl i)} On one side there is the very popular ``expanding space'' picture which claims that it is the 
very fabric of $3$-space that is growing as time passes by, thus giving rise to the observational effects 
like the recession of galaxies. Imported to the local\footnote{By ``local'' we mean  basically the solar 
system scale.} ground, albeit perhaps with different nuances, there is the 
view that such an effect may exist \cite{Sereno:2007tt, Cooperstock:1998ny,Iorio:2013moa,Carrera:2008pi}. 
To put it simply, its effect on the local dynamics\footnote{We do not claim that these authors 
endorse the expanding space picture, their work just being that of examining the consequences of such an assumption at the local 
scale.} of a particle would boil down to an 
additional repulsive acceleration term, a functional of the scale factor $a(t)$ present 
in the FLRW background metric, and of the particle's position. 

{\sl ii)} On the other side, other analysis ignore every fact about the expanding universe, reducing its 
local effect to the presence of a non--vanishing cosmological constant. In doing so one applies 
locally the Schwarszchild de Sitter metric in its static form, leaving no room whatsoever for the 
very idea of any possible effect of the expansion itself \cite{Kagramanova:2006ax}. 

It clearly seems that both pictures can't hold simultaneously. 
We will try in this paper to elaborate in 
favour of what we think is the correct standpoint. It is based on an obvious fact, 
to wit, that when examining the adequacy of a metric in order to describe a certain physical situation, 
one must ensure its consistency with the right hand side (rhs) of Einstein's equations, that is, the matter
energy-momentum tensor\footnote{By ``matter'' we include ordinary 
matter, radiation, dark matter, dark energy --understood as the cosmological constant--, or in general 
whatever source that we put in the rhs of Einstein equations.}. 

Consider for instance the usual layman  question: {\sl if space is expanding, does this means
that my home is expanding?}, followed with the intriguing: 
{\sl but, if my measuring stick is expanding too, how can I measure such an effect 
in the first place?}. If we take a look at the Einstein equations at our local scale, we will 
find an answer to the former question, which in turn makes the latter void of content.
What can one infer from the Einstein equations at our local scale? First and 
foremost: that, except for the possible presence of a cosmological constant, there is no trace 
whatsoever of the homogeneous Hubble flow which sources the FLRW metric\footnote{
We will adopt in the following the obvious simplification to consider the effect of the CBR or 
the neutrino background on the FLRW scale factor negligible, as if the intergalactic vacuum were only 
permeated by the cosmological constant.}. The reason is more than obvious: 
the Hubble flow is the averaged picture of the distribution of matter that only works at much, 
 much larger scales, 
than the local one considered here. And therefore, {\sl the FLRW metric is just a broad-brush, 
coarse-grained, averaged picture of the real metric of spacetime, only apt to describe phenomena 
at the cosmological scale.} Simply we can not continue to use this concept of Hubble flow 
at the local scale and insist on its homogeneity. 
Thus there is no expanding space at all at our, local scale. Take for instance the solar system, 
and adopt as approximately valid the simplifying assumption of spherical symmetry of the matter 
external to it, then we find ourselves in the framework of the Einstein Straus 
approach \cite{EStraus} in which clearly the Hubble flow has no effect --except 
for its cosmological constant component-- on the local system.

 This paper presents a critical assessment of some of these widespread misunderstandings in cosmology, 
for a comprehensible introduction see \cite{tamara}, while we investigate under which circumstances 
local and cosmological physics match. We start by reviewing the de Sitter spacetime and its 
cosmological incarnations, Sec. \ref{dSspacet}. In 
particular we show a change of coordinates to describe the spacetime metric 
surrounding a comoving observer 
in geodesic motion. In Sec. \ref{localeff} we  discuss under which circumstances the dynamics for 
geodesics in a cosmological de Sitter metric is physically equivalent to that induced 
by the dS component in a static Schwarzschild-dS . This brings the opportunity 
to discuss, Sec. \ref{stvsnst}, the role of non-static metrics and the distinguished Schwarzschild-dS 
coordinatization to describe the local scale.  On the other side, at the cosmological scale, we 
show that the only possible source of the energy-momentum tensor to bring a FLRW metric to a static form 
is that of a cosmological constant.
We end, Sec. \ref{reds}, by elaborating on the interpretation of gravitational/cosmological redshifts  
as Doppler effect generalized to General Relativity.

In Appendix \ref{Adopler} we have gathered the construction of an invariant for the Doppler effect 
and its generalization to General Relativity. 
Applications of this generalization to the massive case are discussed in 
 Appendix \ref{massDop}.

%%%%%%%%%%%%%%%%%%%%%%%%%%%%%%%%%%%%%%%%%%%%

\section{ de Sitter spacetime: static and cosmological incarnations}
\label{dSspacet}
%%%%%%%%%%%%%%%%%%%%%%%%%%%%%%%%%%%%%%%%%%%%
Since de Sitter (dS) spacetime  plays a relevant role throughout this paper,  
we review in the sequel its main features. It 
can be defined \cite{Hawking:1973uf} by a $4$-dimensional embedding in a flat, $5$-dimensional 
 Minkowski spacetime, ${\mathcal M}^{(1,4)}$, with coordinates 
 ${\boldsymbol Z}=( x_0,\,x_4,\,x_1,\,x_2,\,x_3)$ 
with Lorentzian metric %with signature $(-,+,+,+,+)$,
\beq
ds^2_{{\mathcal M}^{(1,4)}} = -dx_0^2 +dx_4^2 +dx_1^2+dx_2^2+dx_3^2\,.
\label{metr5d}
\eeq
Then  the dS background is described by the hyperboloid submanifold
\beq{\boldsymbol Z}^2= -x_0^2 +x_4^2 +x_1^2+x_2^2+x_3^2=\frac{1}{H^2}, \quad
H^2:=\frac{\Lambda}{3}\,,
\label{embedd0}
\eeq 
with the cosmological constant $\Lambda >0$. 
One can define coordinates $(T,\,x_1,\,x_2,\,x_3)$ for the dS submanifold by 
\bea
x_0=\frac{\sqrt{1-H^2 R^2}}{H} \sinh (H T)\,,\,\,
x_4=\frac{\sqrt{1-H^2 R^2}}{H} \cosh (H T)\,,\,\, 
 R=\sqrt{x_1^2+x_2^2+x_3^2}\, 
\label{embedd}
\eea
so that (\ref{embedd0}) is satisfied and the metric induced on the quadric by the ambient metric 
(\ref{metr5d}) becomes 
\beq
ds^2 = -(1 - H^2 R^2)\, dT^2 +\frac{1}{ (1 - H^2 R^2)}\, dR^2 + R^2\, d\Omega^2\,,
\quad d\Omega^2= d\theta^2+ \sin^2(\theta)\,d\varphi^2\,.
\label{static-dS}
\eeq
Two remarks are in order: {\sl  i)} 
 these coordinates do not cover the whole 
hyperboloid  (\ref{embedd0}). {\sl ii)} A static observer in this background, $ 
(R,\, \theta,\,\varphi)$ constant, is constantly accelerating. 

Starting from (\ref{static-dS}), we introduce, under the guidance 
of some physical principles, the cosmological, spherically symmetric versions of dS.
The key point is to obtain a radial coordinate
such that the observers comoving with it are time-like geodesics. 
Thus our first task is to find the radial geodesics for the background (\ref{static-dS}). 
In fact we can study a more general case. 
Consider spherically symmetric static metrics of the form
\beq
ds^2 = -f(R)\, dT^2 + \frac{1}{f(R)} dR^2 + R^2 d\Omega^2\,,
\label{staticgen}
\eeq
which include Schwarzschild, dS, AdS, Schwarzschild-dS or Schwarzschild-AdS metrics. 
We look for the equations for the radial time-like geodesics $(T(s), R(s), \theta_0,\varphi_0)$ 
in terms of proper time $s$. The $4$-velocity, $V(s) = (T'(s), R'(s), \theta_0,\varphi_0)$, and the 
proper time condition $V^2=-1$ implies $\d T'(s)^2 = \frac{f(R(s)) + (R'(s))^2}{f(R(s))^2}$. 
Implementing this into the geodesic equation,
$ V^\mu \nabla_\mu V^\rho = \frac{d}{d\,s}V^\rho + V^\mu\Gamma_{\!\mu\nu}^{\rho}\,V^\mu = 0\,,
$
one obtains the radial equation
\beq
\frac{1}{2}\,f'(R(s)) + R''(s)=0\,.
\label{eqf}
\eeq
 Thus for 
(\ref{static-dS}) we have 
\beq
R''(s) -H^2 R(s)=0\quad   \Rightarrow \quad  R'(s)^2 -H^2 R(s)^2 = C \,,
\label{radeq}
\eeq 
being $C$ a constant. Time-like radial geodesics are classified according to the sign of this 
constant. Assuming conventional initial conditions at $s=0$, the different solutions to (\ref{radeq}) are:
\bea
C=0&:&\qquad R(s) = r\, {\rm e}^{H s}, \hspace{24mm}
 {\rm e}^{-2 H\, T(s)} ={\rm e}^{- 2 H s}-H^2 r^2 \,.\nonumber\\
C>0&:&\qquad R(s) = r \sinh(H s), \quad \tanh (H\, T(s))=\sqrt{1+H^2 r^2} \tanh (H s)\,.\nonumber\\
C<0&:&\qquad R(s) = r \cosh(H s), \quad \tanh (H\, T(s))=\frac{\tanh (H s)}{\sqrt{1-H^2 r^2}}\,,
\label{3sols}
\eea
with $r$ constant and T(s) given in implicit form. 
For vanishing $C$ there exist in addition the solution $R(s) = r {\rm e}^{- H s}$ 
which represents geodesics moving inwards instead 
  of outwards. Notice that, from the viewpoint of the static observers in the background 
  (\ref{static-dS}), the  constant $C$ is, in Newtonian language, 
  proportional to the conserved energy of the corresponding geodesic motion.

Eqs. (\ref{3sols}) suggest three different changes of coordinates
based on two physical requirements: {\sl i)} The radial geodesics become comoving 
in the new coordinate system. It is then obvious that the parameter $r$ 
is the natural choice for the new radial coordinate. {\sl ii)}
We require the proper time $s$ of these comoving geodesics, to become the new time 
coordinate $t$.

On the whole we are proposing three different 
changes of coordinates $(T,R)\rightarrow (t,r)$ that can be read from (\ref{3sols}) by writing
$t$ instead of $s$ and $R(t, r)$ and $T(t, r)$ in place of $R(s),\, T(s)$. 
The metric (\ref{static-dS}) is written, in the new coordinates, as
\bea
C=0&:&\qquad  ds^2 = -dt^2 + {\rm e}^{2 H\,t}(dr^2 + r^2 d\Omega^2)   \,.\nonumber\\
C>0&:&\qquad   ds^2 = -dt^2 + \sinh(H\,t)^2\Big(\frac{1}{1+ H^2 r^2} dr^2 + 
r^2 d\Omega^2\Big) \,,\quad C= H^2 r^2\,.\nonumber\\
C<0&:&\qquad  ds^2 = -dt^2 + \cosh(H\,t)^2\Big(\frac{1}{1- H^2 r^2} dr^2 + 
r^2 d\Omega^2\Big)  \,,\quad C= -H^2 r^2\,,
\label{3metrics}
\eea
which are the well known cosmological dS metrics with the flat, hyperbolic and spherical slices, 
respectively. 
The physical arguments used to produce these changes of coordinates will be a guiding principle  in the discussions in the next section.
 
%%%%%%%%%%%%%%%%%%%%%%%%%%%%%%%%%%%%%%%%%%%%
\section{ Local effects of the expansion  }
\label{localeff}
%%%%%%%%%%%%%%%%%%%%%%%%%%%%%%%%%%%%%%%%%%%%

Let us consider the general FLRW cosmology, with metric
\beq
ds^2 = -dt^2 + a(t)^2\Big(\frac{1}{1- \sigma r^2} dr^2 + r^2 d\Omega^2\Big)\,,
\label{gencosm}
\eeq
being $\sigma$  a constant.
Although, as mentioned above,  this metric is a valid description of spacetime at the cosmological scale 
 let's us elaborate on the outcomes assuming 
for a while its correctness at the local scale.
 It is straightforward to quantify the local effect of being in such cosmological background. 
One should find a suitable  description of the time evolving physical distance between a comoving 
observer at radial coordinate $r$ and the center $r=0$. Without entering into fine details, is is 
clear that a function like $R(t)= a(t)\,r$ accomplishes this goal\footnote{ It is worth noticing that, after checking with  (\ref{3metrics}), the trajectories 
$R(s)$ in (\ref{3sols}) can be written in this form. }.
In fact one realises that for 
$\sigma=0$, it gives the distance from the location $(r,\,\theta,\,\varphi)$ to the origin $r=0$, 
obtained with the metric induced from (\ref{gencosm}) on the equal time hypersurface\footnote{For 
$\sigma\neq 0$ that would not be exactly the distance, but it will remain a good approximation 
as long as $\sigma r^2 << 1$.}. Clearly this new time dependent radial variable satisfies 
\cite{Carrera:2008pi,romano12,Faraoni:2007es},
\beq
 R''(t)=  a''(t)\,r =\frac{ a''(t)}{a(t)}\,R(t)\,,
\label{uddot}
\eeq
and thus it makes the case that  the effect of the expansion on the local dynamics, 
either at the scale of the planetary orbits or at that of the electronic orbits of an atom, 
is just a repulsive\footnote{ As long as  $a''(t)>0$.} radial acceleration, proportional to 
 the distance $R(t)$ and to the time-time component of the Ricci tensor $\d\frac{ a''(t)}{a(t)}$. In fact, 
the more complete discussion in \cite{Cooperstock:1998ny}, with the use of Fermi coordinates, 
leads to the same results. Thus one concludes that, {\sl if such FLRW (\ref{gencosm}) 
models were valid at the local scale, then (\ref{uddot}) would capture the effect of the expansion}.
But, as argued above, except for the possible presence of a cosmological 
constant, there is no trace of the Hubble flow at the local scale --just check the rhs of Einstein 
equations. This would make the general derivation of (\ref{uddot}) just an interesting and 
valuable academic exercise, except that it turns to be right in a 
qualified and restricted sense: a contribution to the rhs of (\ref{uddot}) remains 
at the local scale when the Hubble flow includes a cosmological constant component. In such case, the 
cosmological constant will remain as the only local effect of the Hubble flow, because it is 
present {\sl at all scales}. In addition to that, in the case that $a(t)$ 
corresponds to dS or AdS spacetimes, and in view of (\ref{3sols}) and (\ref{3metrics}),  
the variable $R$ in (\ref{uddot}) is nothing but the radial variable in the static 
coordinatization (\ref{static-dS}). 

The cosmological dS spacetimes 
have been studied in Sec. \ref{dSspacet} and the procedure to deal with AdS spacetime is analogous. 
In all these cases the same relation holds, 
$\frac{ a''(t)}{a(t)} = \frac{\Lambda}{3} $, 
with $\Lambda$ positive for dS and negative for AdS. Thus, either for dS or AdS, (\ref{uddot}) 
becomes 
\beq
R''(t) =  \frac{\Lambda}{3}R(t)\,.
\label{ufordS}
\eeq

Up to here we have analysed the effects of the cosmological constant in the framework of cosmological
 FLRW models. Now let us turn our attention to the approach \cite{Kagramanova:2006ax} where
the presence of a cosmological constant  at the solar system scale is examined by using
a static version of the Schwarzschild-dS metric \cite{Kottler}, in which (\ref{staticgen}) is materialized with
\beq\displaystyle f(R) = 1-\frac{2\,M}{R}- \frac{\Lambda}{3}\,R^2
\label{static-Schw-dS}\,.
\eeq
The radial geodesic equation is obtained from (\ref{eqf}),
\beq
R''(s) =- \frac{M}{R(s)^2}+\frac{\Lambda}{3}\,R(s)  \,,
\label{uforSchwdS}
\eeq
which contains a Newtonian leading term in addition to
the acceleration coming from the presence of $\Lambda$, which we isolate,
\beq
R''(s) = \frac{\Lambda}{3}\,R(s) \,.
\label{uforstatdS}
\eeq
It is centrifugal for dS and centripetal for AdS. 

Results in (\ref{ufordS}) 
and (\ref{uforstatdS}) match, but this could be just a formal coincidence because of the notation. 
So we must look closely at the physical meaning attached to (\ref{ufordS}) and (\ref{uforstatdS}). 
This has been already pursued in Sec. \ref{dSspacet}, where it was shown that the time
coordinate  $t$ in the cosmological dS settings (\ref{3metrics}) is just the proper time $s$ of 
the radial geodesics of (\ref{static-dS}). 
We conclude therefore that (\ref{ufordS}) and (\ref{uforstatdS}) are equations with identical content. 

The first lesson we extract form this analysis is that equation (\ref{uddot})  
is indeed correct if one restricts its application to the only acceptable case, that
of a Hubble flow including a cosmological constant component.

The second lesson is that, at the local scale, the effect of the Hubble flow is completely 
captured with a static metric, of which (\ref{static-Schw-dS}) is a good example, and there is no 
need to implement a time dependent metric. 

We elaborate slightly more in this second point in the following.

%%%%%%%%%%%%%%%%%%%%%%%%%%%%%%%%%%%%%%%%%%%%
 \section{Static vs. non--static metrics}
 \label{stvsnst}
%%%%%%%%%%%%%%%%%%%%%%%%%%%%%%%%%%%%%%%%%%%%

%%%%%%%%%%%%%%%%%%%%%%%%%%%%%%%%%%%%%%%%%%%%
 \subsection{A tale at the local scale}
\label{statvsn}
%%%%%%%%%%%%%%%%%%%%%%%%%%%%%%%%%%%%%%%%%%%%
 
Let us take the two versions of dS spacetime, ({\sl a}) the static (\ref{static-dS}) 
and ({\sl b}) the expanding (\ref{3metrics}). 
While static observers in the former do not observe time evolution 
in their spacetime,  static observers in the latter do notice an expanding universe. 
The way out to this apparent paradox is that static observers in ({\sl a})  and ({\sl b}) do not 
share the same physical properties. As a matter of fact the static observers in ({\sl a}) are 
{\sl constantly accelerating}, whereas the static observers in ({\sl b}) are {\sl geodesics}.
What is the most
convenient coordinatization in order to facilitate the quantitative description of physical measurements?
To find the answer it may be helpful if we consider Schwarszchild spacetime 
and examine some coordinate systems available to describe it.

Before getting into details, we may establish two consecutive stages as regards the 
connection between 
the mathematical coordinates, quite arbitrary because of diffeomorphism invariance, and the 
measuring devices.

\begin{enumerate}
\item[(I)] In the first stage, observables in General Relativity\footnote{Diffeomorphism 
invariant objects.} (GR) are defined 
through a gauge fixing {\sl i.e.} after a common choice, made by all the observers, of a 
coordinatization \cite{Thiemann:2004wk,Dittrich:2005kc,Pons:2009cz}. 

\item[(II)] The second stage is the trickiest one: to connect the 
coordinates with the physical measuring devices. 
That is, given the  observational features that the particular users are focusing on, 
to optimize its mathematization through the adoption of coordinate descriptions suited to their 
measuring devices. 
\end{enumerate}

Now consider the solar system scale. 
As regards considerations of static versus non-static descriptions, we must
point out that the success of the standard coordinatization for the Schwarszchild metric makes such 
choice the preferable option. As a consequence of Birkoff's theorem 
the correccions to the solar system 
Newton dynamics are perturbative terms, and these obtain maximal simplicity --just a single term-- 
within the standard Schwarszchild coordinatization \cite{hakmann}, whereas for instance using isotropic coordinates 
--also static-- the corrections are distributed among several terms.
As said, static observers in the standard form for Schwarzschild metric are constantly 
accelerating. In fact, and probably with some degree of retrospection, it is intuitively clear that 
coordinates for which a static observer is constantly accelerating have a good physical content to 
describe, precisely, geodesic motion. The reason is inspired in Newtonian physics, in which ideal 
static observers, placed around the sun, have an internal experience of constant acceleration 
(to keep them at rest).

By passing let us mention that the authors of \cite{NG} have provided with an intrinsic definition of 
distance, disconnected from Newtonian physics, showing the physical preference of the standard form 
of the Schwarzschild metric in order to describe the geodesics associated to planetary orbits 
in general relativity.

If in addition to the Schwarszchild picture we include the presence of a cosmological constant, 
we may consider Schwarzschild-dS in the forms (a) static, (\ref{static-Schw-dS}), and (b) 
time dependent\footnote{This cosmological Schwarzschild-dS could be constructed in principle by the 
procedure used in Sec. \ref{dSspacet}.} . 
According to the observations derived from the previous analysis in Sec. \ref{localeff}, 
in which we 
saw that the additional acceleration induced by the cosmological constant is already accounted for 
in the static version of dS, 
it seems quite clear that the static description (a) is the 
candidate to be the most convenient one in order to describe physical phenomena at the local scale.

We have considered hitherto the solar system scale up to the galactic scale. In addressing
 the cosmological 
scale, we find that there is a drastic limitation for the FLRW metrics that admit static 
versions. This is the subject of the next section. 
 
%%%%%%%%%%%%%%%%%%%%%%%%%%%%%%%%%%%%%%%%%%%%
 \subsection{A tale at the cosmological scale}
\label{canwe}
%%%%%%%%%%%%%%%%%%%%%%%%%%%%%%%%%%%%%%%%%%%%

Let us consider the general FLRW cosmology (\ref{gencosm}). We will answer the question as to 
whether and when such background admits a static metric\footnote{The answer was already given in 
\cite{florides} but we take a completely different approach, which we believe is more advantageous and  direct.}.
With this aim, since we know that comoving observers in a static metric are 
constantly accelerating,
we should look for the radial trajectories of the constantly accelerated observers. One 
obtains a first change of variables $r(t,R)$ such that, after a second change  
$t(T,R)$ we end up with an static metric, being $R$ its new  radial coordinate. Examining the metric 
coefficient for the two-sphere it is clear that $a(t)\,r(t,R)$ must be a function of $R$ only, 
the simplest choice being $R$ itself. Thus
one can identify $\displaystyle r(t,R)=\frac{R}{a(t)}$. The requirement of being static translates to that of a constantly accelerated 
observer for constant $R$.
As a consequence  one must consider the trajectory
$X=\{t,\frac{R}{a(t)}, \theta_0,\varphi_0\}\,.$ Its
$4$-velocity wrt proper time $s$ reads 
\bea
V= \Big\{\frac{1}{\sqrt{\frac{R^2 a'(t)^2}{\sigma  R^2-a(t)^2}+1}},
-\frac{R\, a'(t)}{a(t)^2 \sqrt{\frac{R^2 a'(t)^2}{\sigma R^2-a(t)^2}+1}},0,0\Big\}\,,\qquad 
 V^2= -1\,,
\eea
and the acceleration is computed as 
$A^\mu= V^\rho\nabla_\rho V^\mu = \frac{d}{d\,s}V^\mu + V^\rho \Gamma_{\rho\nu}^\mu V^\nu$.
 %, that is, $ \frac{d}{d\,s}V^\mu = V^\rho \partial_\rho V^\mu$. 
Next we should compute the Jerk, defined as the Fermi-Walker covariant derivative of the $4$-velocity  
\cite{Russo:2009yd,Pons:2018lnw} and implement the equation ${\rm Jerk}=0$, which describes the constantly 
accelerated trajectory. Due to the symmetry of our setting, one can check that the 
vanishing of the Jerk is equivalent to the constancy of the norm of the acceleration. One obtains
\beq
|A|^2= \frac{1}{a(t)^2} \frac{E_1^2}{E_2^3} \,, \quad {\rm with}\quad
\left\{ \begin{matrix}
E_1 = R\, a(t)\, a''(t)\Big(\sigma\,  R^2-a(t)^2\Big)+R^3 a'(t)^2 \Big(a'(t)^2+\sigma \Big)\,, \\
E_2 = a(t)^2-R^2 \Big(a'(t)^2+\sigma \Big)\,.
\end{matrix}
\right.
\label{normA}
\eeq
Requiring $|A|=$ constant will in general give solutions for $a(t)$ containing $R$ dependences. 
This translates contrariwise to our starting point, (\ref{gencosm}), into $r$ dependences in 
$a(t)$\footnote{Notice however that $a(t)$ may indeed depend on $\sigma$, 
which is a parameter already present in the metric.}.
The only way in order for the equation $|A|=$ constant not to display an $R$ dependence 
for its solution $a(t)$, is that the coefficients of different powers of $R$ in the numerator and denominator of 
(\ref{normA}) must sustain a relation of 
proportionality. With these considerations we set up the following relation
\beq
\frac{{\rm Coefficient}(E_1, R^3)}{{\rm Coefficient}(E_1, R)}= 
\frac{{\rm Coefficient}(E_2, R^2)}{{\rm Coefficient}(E_2, R^0)}\,,
\eeq
which yields the condition
\beq
a(t)\,a''(t) -a'(t)^2 - \sigma=0 \,.
\label{req}
\eeq
 We recognize in the above equation the condition for (\ref{gencosm}) to be 
a maximally symmetric spacetime \cite{Weinberg:1972kfs}\footnote{ We can check that the solutions to (\ref{req}), see below,  
satisfy the requirement $|A|=$ constant.}. 
In addition (\ref{req}) is the EOM  
for the Lagrangian $\displaystyle {\cal L_\sigma} =\frac{a'(t)^2-\sigma}{a(t)^2}$, 
which associated Hamiltonian, 
$\displaystyle {\cal H_\sigma} =\d  \frac{a'(t)^2+\sigma}{a(t)^2}$,
is a constant of motion that is nothing but $\d \frac{\Lambda}{3}$, 
being $\Lambda$ the cosmological constant. Using this fact (\ref{req}) boils down to\footnote{ Notice that this equation is derived form the 
Lagrangian $\d {\cal L}_\Lambda = a'(t)^2+\frac{\Lambda}{3}\,a(t)^2 $, 
in which case $\sigma$ appears as the new Hamiltonian constant of motion. Thus equations (\ref{req}) 
and (\ref{req2}) are equivalent, and so they are their parent Lagrangians 
${\cal L_\sigma}$ and ${\cal L}_\Lambda $.} 
\beq
a''(t) -\frac{\Lambda}{3}\,a(t) =0\,.
\label{req2}
\eeq
Its solution\footnote{ We skip the trivial solution $a(t)=$ constant, $\sigma=\Lambda=0$, which
is Minkowski spacetime in standard coordinatization.} 
contains three different cases, depending on the sign of $\Lambda$:
\begin{enumerate}
\item[(I)]
The case $\Lambda >0$ is dS spacetime which
has already been dealt with in Sec. \ref{dSspacet}. 
\item[(II)] The vanishing $\Lambda$ case is Minkowski spacetime in Milne coordinatization \cite{Milne}\footnote{The change of variables from Minkowski, $ds^2=-dT^2+dR^2+R^2 d\Omega^2$, to Milne coordinates is given by $T(t,r)=\sqrt{1+H^2 r^2}\, t\,\,,\,R(t,r)=HrT$.}
\beq
ds^2 = -dt^2 + H^2t^2\Big(\frac{1}{1+ H^2 r^2} dr^2 + r^2 d\Omega^2\Big)\,.
\label{Milne}
\eeq
\item[(III)] The case $\Lambda< 0$ is AdS spacetime. It needs $\sigma<0$ and its equal time 
  slices are $3$-hyperboloids
 \beq
  ds^2 = -dt^2 + (\sin(H\,t))^2\Big(\frac{1}{1+ H^2 r^2} dr^2 + r^2 d\Omega^2\Big)\,,
  \quad  H^2:=-\frac{\Lambda}{3}=-\sigma. 
 \label{AdS}
 \eeq
\end{enumerate}
  
All these solutions share the same description with a static metric,
  \beq
ds^2=-(1-\frac{\Lambda}{3} R^2)\, dT^2 + \frac{1}{1-\frac{\Lambda}{3} R^2}dR^2  
+ R^2(d\theta^2 + (\sin\theta)^2 d\varphi^2)\,.
\label{(A)dSstatic}
\eeq

To summarize, the only source of matter-energy compatible 
with bringing a given FLRW cosmology to a static metric description is the cosmological constant.

%%%%%%%%%%%%%%%%%%%%%%%%%%%%%%%%%%%%%%%%%%%%%%%%
\section{Demystifying some folklore}
\label{reds}
%%%%%%%%%%%%%%%%%%%%%%%%%%%%%%%%%%%%%%%%%%%%%%%%%

\subsection{Is the $3$-space expanding?}

Let us give a tentative definition, within the FLRW models, of what is meant by expanding $3$-space: 
 it is the idea, or belief, that there is a physical process of some sort 
--acting perhaps at an ultra-micoscopic scale-- that is producing the homogeneous growing of the 
equal-time hypersurfaces of the background (\ref{gencosm}), as dictated by the scale factor $a(t)$. 

Take a family of non--interacting test particles following radial
time-like geodesics inside a Minkowski 
spacetime. Suppose we adopt Milne's form (\ref{Milne}) 
for the metric. Hence test particles are comoving observers with scale factor $H\,t$
describing the time-increasing separation between them.
Should one conclude from this picture that the $3$-space is expanding? \cite{Peacock:2008uc}. 
Obviously such notion is a pure artifact of the chosen coordinatization.
On the other hand, there is a strong observational evidence that the galaxies are 
moving apart from each other.  
Provided the intergalactic void is not so different from the strict void described 
by (\ref{Milne})\footnote{Of course the former is crossed by electromagnetic radiation, neutrinos... 
and feels the pervading presence of the tiny cosmological constant.}, 
should we believe that the $3$-space in our universe
expands whereas that in (\ref{Milne}) does not?. We find neither compelling reason nor need to believe 
in the expansion of $3$-space as defined above.
This picture is not sensible and receives the final blow when realizing
 that if taken seriously, 
then one is bound to accept the absurd 
consequence that this growing process holds at the local scale, for which there is no basis at all 
when one looks at the rhs of Einstein's equations.

\subsection{Cosmological and gravitational redshift vs. Doppler effect}

Take two test observers in Minkowski spacetime, A and B, simultaneously departing from the origin 
of coordinates at $T=0$ and traveling radially in different directions. They correspond, see (\ref{Milne}), to comoving observers located at 
fixed $\{r_a, \theta_a, \varphi_a\}$ and $\{r_b, \theta_b, \varphi_b\}$ in the Milne coordinatization. 
One can compute the redshift of a photon emitted by A and detected by B using the Special 
Relativity (SR) formulas for the Doppler effect, 
when sender and receiver are in different
inertial reference systems. Although this
 redshift can also be computed as a byproduct of the "expansion of 
space" picture, it is clearly nothing else than a Doppler effect. 

The Doppler effect as introduced above in the framework of SR can be generalized to GR as relating the frequencies of the emitted and received photons by 
arbitrary sources and observers. 
This was done long time ago, first in \cite{schrodinger,J.L.Synge:1960zz} and later re-elaborated 
\cite{Narlikar,Lewis:2016gzx}. 
Once this generalization of the Doppler effect to curved spacetime is at our disposal, 
all these considerations on a redshift that is no longer Doppler but only cosmological, 
 or gravitational, become meaningless. To avoid misunderstandings: 
It is not wrong to talk on cosmological redshift, 
or gravitational redshift, what is wrong is to claim that 
they are something conceptually different from the Doppler redshift, now understood as an 
extension to GR of the SR effect. This extension is reviewed in the Appendix \ref{Adopler}.

In the next two subsections we examine the cosmological and gravitational redshifts within the framework 
of this generalization. It is worth noticing in these derivations the relevance of identifying an affine parameter 
for the photon trajectory. 
In Appendix B we discuss the massive particle case, and show that we recover in the massless limit the results 
shown below, thus becoming an alternative derivation of the frequency shifts for the photon, only relying on the
notion of proper time.

%%%%%%%%%%%%%%%%%%%%%%%%%%%%%%%%%%%%%%%%%%%%%
\subsubsection{ The cosmological redshift as a General Relativity Doppler effect}
%%%%%%%%%%%%%%%%%%%%%%%%%%%%%%%%%%%%%%%%%%%%%

As is elaborated in depth in App. \ref{Adopler} the spectral shift, $z$, relates
the  frequency $\nu_s$ of the photon as seen by 
the Source
at the time of emission, to the frequency $\nu_o$ of the photon as seen by  the 
Observer at the time of reception\footnote{In both cases the frequency is proportional to the 
kinetic energy.}. This relation can be cast  in an invariant way as (\ref{easydopplergen2}) 
\beq
1+z= \frac{\nu_s}{\nu_o}= \frac{E_s}{E_o}= 
\frac{V_s(0)\cdot U(0)}{V_o(t)\cdot U(t)}\,,
\label{1+z}
\eeq
with $t=0$ at the emission event.
The Source and Observer's velocities, $V_s(t)\ {\rm and}\ V_o(t)$, 
computed wrt to  their
respective proper time, are 
 parametrized here by the coordinate time in the spacetime. While $U(t)$ is the velocity of 
the photon, still parametrized by the coordinate time, but computed 
wrt an affine parameter\footnote{Lacking
of the concept of proper time for the massless particle, this is the {\sl only 
way} to ensure that the quotient in (\ref{1+z}) is indeed an invariant. }.

In the sequel we shall apply the previous expression, (\ref{1+z}), to verify 
that it describes the cosmological redshift for the FLRW, (\ref{gencosm}). 
We consider both the Source and the Observer as comoving in the 
Hubble flow. The Source is located at $r=0$, in this way 
the spherical symmetry imposes that all geodesics passing through
$r=0$ must be radial. In addition, the maximal symmetry of the equal time slices allows that any 
non--radial geodesic become automatically included in the analysis through
a change of coordinates in the equal-time slices. 
With the aforementioned conditions, the photon trajectory is given by
$
X(t)= \Big(t, r(t),\theta_0,\phi_0\Big)\,,
$
with r(t)=0 and velocity wrt coordinate time
$\d \frac{d\,X}{d\,t} = (1, r'(t),0,0)$. The velocity wrt
the affine parameter $s$ is,
\beq 
U(t) = \frac{d\,t}{d\,s}\, \frac{d\,X}{d\,t} \equiv h(t) \Big(1, r'(t),0,0\Big)\,,   \quad  
\frac{d\,t}{d\,s} 
:=h(t(s))\,,
\eeq
where $h$ is as yet an unknown function. The equation for $h(t)$ is found by inserting 
$U(t)$ into the geodesic equation,
\beq
\frac{d\,U^\mu}{d\,s} + \Gamma^\mu_{\ \nu\rho}\,U^\nu U^\rho =
h(t)\, \frac{d\,U^\mu}{d\,t} + \Gamma^\mu_{\ \nu\rho}\,U^\nu U^\rho =0\,,
\label{geodeq}
\eeq
from which we obtain a differential equation for $h(t)$, $\displaystyle h(t)\, a'(t)+a(t)\,  h'(t)=0$, with 
solution\footnote{The arbitrary constant factor of the solution will be discussed later on.}
\beq
h(t)=\frac{1}{a(t)}\,.
\label{affine}
\eeq
Imposing the light-like condition $\d \Big(\frac{d\,X}{d\,t}\Big)^2=0$, which becomes
$\displaystyle \frac{a(t)^2 r'(t)^2}{1-\sigma\,  r(t)^2}=1$\,,
one determines the trajectory\footnote{Note that $\displaystyle \int_0^t \frac{1}{a(\tau)} \, d\tau$ 
s the conformal time $\eta$, with 
$d t= a(t(\eta))d\eta$. } 
\beq
r(t)= \frac{1}{\sqrt{\sigma }}\sin \left(\sqrt{\sigma } \int_0^t \frac{1}{a(\tau)} \, d\tau\right)\,
\label{r-of-t}
\eeq
which holds for both positive or negative $\sigma$ and also in the limit of vanishing $\sigma$. 
Thus we end up with
\beq
U(t)= \Big(\frac{1}{a(t)},
\frac{\cos \left(\sqrt{\sigma } \int_0^t \frac{1}{a(\tau)} \, d\tau\right)}{a(t)^2},0,0\Big)\,.
\label{U-of-t}
\eeq
The comoving Source at the time of emission 
$t=0$ has 
proper velocity $V_s(0) = (1,0,0,0)$, and thus
$\displaystyle V_s(0)\cdot U(0) = \frac{1}{a(0)}$. On the other hand, the comoving Observer at the 
time $t$ of reception has proper velocity $V_o(t) = (1,0,0,0)$, hence 
$\displaystyle V_o(t)\cdot U(t) = \frac{1}{a(t)}$. All in all, from  (\ref{1+z}) we get
\beq
1+z= \frac{\nu_s}{\nu_o}=
\frac{V_s(0)\cdot U(0)}{V_o(t)\cdot U(t)} = \frac{a(t)}{a(0)}\,,
\label{1+z massless}
\eeq
which is the standard formula for the cosmological redshift. 

This result shows that the cosmological redshift 
is just the manifestation of the Doppler effect, once extended from SR to GR.
Nothing more, nothing less. Of course one can use the rubber balloon 
picture \cite{Edd33} as a metaphor, but to claim that 
such a picture is necessary in the sense of our tentative definition given above
is a mistake.

%%%%%%%%%%%%%%%%%%%%%%%%%%%%%%%%%%%%%%%%%%%%%
\subsubsection{The gravitational redshift as a General Relativity Doppler effect}
%%%%%%%%%%%%%%%%%%%%%%%%%%%%%%%%%%%%%%%%%%%%%
\label{gravredshift}
Once the Doppler effect has been properly extended to GR, we can conclude that 
cosmological and gravitational redshifts have a 
common origin. This result was already derived in \cite{schrodinger,J.L.Synge:1960zz,Narlikar} 
and in the 
following we make a detailed treatment for the general Schwarszchild-dS spacetime.

Consider the general metric (\ref{staticgen})
%\beq
%ds^2 = -f(R)\, dT^2 + \frac{1}{f(R)} dR^2 + R^2 d\Omega^2\,.
%\label{staticgen2}
%\eeq
with $f(R)$ given in (\ref{static-Schw-dS}). 
We are interested in the radial emission of a photon from a location 
$(R_0,\theta_0,\phi_0)$ and its latter detection along the same radial line at 
 $(R_1,\theta_0,\phi_0)$ with $R_1>R_0$. 
The Source (Observer) position and velocity wrt proper time\footnote{ Thus normalized to $V^2=-1$.} 
are 
\bea
\label{Vo}
\text{Source}:\left\{
\begin{matrix}
X_s=(T,R_0,\theta_0,\phi_0)\\
\,V_s=(\frac{1}{\sqrt{f(R_0) }},0,0,0)\end{matrix}
\right.\,,
\quad
\text{Observer}:\left\{
\begin{matrix}
X_o=(T,R_1,\theta_0,\phi_0)\\
\,V_{o} = (\frac{1}{\sqrt{ f(R_1)}},0,0,0)
\end{matrix}
\right.\,.
\eea
While the photon trajectory\footnote{Not described yet with the necessary affine parameter.} is 
$X_{ph}=(T,R(T),\theta_0,\phi_0)$, with $R(0)=R_0\,\,{\rm and}\,\, R(T)=R_1$. Its
$4$-velocity, $(1,R'(T), 0, 0)$, being a null vector determines the equation $
R'(T)=f(R(T))\,
$
and wrt proper time boils down to $U(T) =h(T) (1,R'(T), 0, 0)$, with $h(T)=\frac{d T }{d s}$ 
and $s$ an affine parameter. Similarly to the previous case, 
the geodesic condition fixes $h(T)$ and $U(T)$ becomes
\beq
U(T) =  (\frac{1}{f(R(T))},1, 0, 0)\,.
\label{Ugrav}
\eeq
At any $T>0$, with $R(T)=R_1$, the scalar products are found to be  
$$
V_{o}(T) \cdot U(T) =-\frac{1}{\sqrt{f(R_1)}}, \quad 
V_s \cdot U(0) =-\frac{1}{\sqrt{f(R_0)}}\,.
$$
Adapting (\ref{1+z massless}) to the case at hand
\beq
z=
\frac{V_s\cdot U(0)}{V_{o}(T)\cdot U(T)}-1 = \sqrt{\frac{f(R_1)}{f(R_0)}}-1\,,
\label{zgrav}
\eeq
which is the usual formula for the gravitational redshift for radial photons in the Swcharzschild 
metric. 
Here it is also valid for Schwarzschild-dS or Schwarzschild-AdS backgrounds.

Notice that contrariwise to the previous discussion on the cosmological 
redshift the Source and Observer are no longer geodesics but constantly accelerating.
One can 
nevertheless consider the Source as belonging to a radial geodesics which happens to be at $R_0$ 
when $t=0$ and such that $R'(0)=0$, and the same 
can be done at time $t$ for the Observer.  The GR Doppler formula still holds 
because the quotient of the scalar products in  (\ref{1+z massless}) is
always an invariant regardless of the fact that Source and Observer be geodesics or not.
In addition, once the Doppler effect has been extended to GR, the curvature of spacetime contributes to this effect, making it detectable even in cases where Source and Observer are at rest\footnote{This  is a coordinate dependent statement.}.

%%%%%%%%%%%%%%%%%%%%%%%%%%%%%%%%%%%%%%%%%%%%
\section{ Concluding remarks  }
\label{cr}
%%%%%%%%%%%%%%%%%%%%%%%%%%%%%%%%%%%%%%%%%%%%

In the previous pages we have focused in relating local and cosmological physics:
{\sl i)} On one side we have matched the radial EOM of a FLRW-dS model with the contribution from a 
cosmological constant of an static Schwarzschild-dS, providing evidence that both models describe the 
same dynamics. In this way, some approaches in the literature that until now seemed to be
incompatible can be reconciled when applied to plausible physical scenarios.
{\sl ii)} On the other side we reviewed an old, but not yet as popular as it deserves, unified presentation 
of the GR Doppler effect, with a single formula encompassing all circumstances. It's common theme being 
that of energy gain or loss for a particle, either massive or massless, in geodesic motion from the Source 
to the Observer.
Under this common theme, all energy shifts, including the cosmological and gravitational ones, appear as 
particular cases of this GR Doppler effect.  Our presentation has the novelty to include in the same
framework the massless as well as the massive case, showing how to retrieve the former from the latter 
by taking the appropriate limit. In this sense, the role of the affine parameter in the massless case can be 
circumvented. We claim that the matching of an invariant at the local and 
cosmological scale provides an unambiguous and unique way to relate observables at both scales.

\subsection*{Acknowledgements}

J.~M.~P.~ is partially supported by research contracts FPA2016-76005-C2-1-P, PID2019-105614GB-C21
(Ministerio de Econom\'ia y Competitividad) and 2017SGR929 (Generalitat de Catalunya).
P.~T.~ is partially supported by grant PID2019-105614GB-C22.

\appendix

%%%%%%%%%%%%%%%%%%%%%%%%%%%%%%%%%%%%%%%%%%%%%
\section*{Appendices}

\section{The Doppler effect as energy gain or energy loss}
\label{Adopler}

\subsection{From Special Relativity...}
The Doppler spectral shift of light in the framework of SR can be understood as a change of the 
energy of a traveling 
photon between emission and reception\footnote{ Synge showed in \cite{Synge:1935zza}
the proportionality of energy and frequency for a photon independently of quantum mechanical 
considerations.}. 
We will skip the usual derivation of this effect and 
examine what is basically equivalent: the energy 
gain/loss of a particle in free motion from the Source to the Observer. This is more general 
that just Doppler, because it includes the massive case.

We consider the inertial reference system of the Observer, placed at the origin of spatial coordinates, 
whereas the Source and the massive particle are moving with respect to it at 
speeds $\vec v$\,, $\vec u$ respectively. Their respective $4$-velocities wrt proper time are
\beq
 V_o=(1,\vec 0)\,,\qquad V_s=(\frac{1}{\sqrt{1-v^2}},\frac{\vec v}{\sqrt{1-v^2}})\,,\qquad 
U=(\frac{1}{\sqrt{1-u^2}},\frac{\vec u}{\sqrt{1-u^2}})\,,
\eeq
with $v = |\vec v|,\ u = |\vec u|$. It is assumed that the particle intersects the Source 
and Observer trajectories at different points in Minkowski spacetime.

If we set the mass of the particle to $m=1$ its energy can be expressed in terms of an invariant form from both rest systems, Observer and Source
\beq
E_o=  -V_o\cdot U=\frac{1}{\sqrt{1-u^2}}\,,\quad
E_s=  -V_s\cdot U=\frac{1-\vec v\,\vec u}{\sqrt{(1-v^2)(1-u^2)}}\,,
\eeq
and therefore, the invariant ratio of energies, $E_s/E_o$, for the massive particle is
\beq
\frac{E_s}{E_o}\ ({\rm massive})= \frac{1-\vec v\,\vec u}{\sqrt{1-v^2}}=
\frac{1-v\, u\,\cos\alpha}{\sqrt{1-v^2}} \,.
\label{easydoppler}
\eeq
In the massless limit, $u\to 1$, we get the standard formula for the energy 
shift of the photon\footnote{Obviously the same result is obtained by using directly for the massless particle the velocity
$U=(1,\vec\omega)$ with  $|\vec\omega|=1$\,.},
\beq
\frac{E_s}{E_o}\ ({\rm massless}) =\frac{1-v\,\cos\alpha}{\sqrt{1-v^2}},
\label{easydopplerless}
\eeq
which, for $\alpha=\pi$, gives the usual longitudinal Doppler redshift when the motions of 
Source and Observer are aligned and in opposite directions.

Summing up, in the SR framework, the Doppler effect or in general, 
the ratio of the particle's energy seen from the
Source rest frame to the particle's energy seen from the
Observer rest frame\footnote{Be the particle either massive or massless.} 
is always described by the invariant
\beq
\frac{E_s}{E_o}= \frac{V_s\cdot U_s}{V_o\cdot U_o} \,,
\label{easydopplergen}
\eeq
with $U_s = U_o = U$ in this case.

%%%%%%%%%%%%%%%%%%%%%%%%%%%%%%%%%%%%%%%%%%%%%
\subsection{...to General Relativity}
%%%%%%%%%%%%%%%%%%%%%%%%%%%%%%%%%%%%%%%%%%%%%

There are many definitions within
the SR framework that can be extended to GR. 
Take for instance the geodesic motion, which is extended to GR by basically 
replacing the ordinary derivative for the covariant one. Or the concept of the constantly 
accelerated observer, that can be brought to GR by 
keeping the requirement of constancy \cite{Pons:2018lnw,Friedman:2016tye,FR17} for some 
curvature scalars that generalize the Frenet-Serret formalism \cite{Letaw:1980yv}. 
These cases bear in common that only the point and its neighborhood in a world line trajectory 
are necessary ingredients. 
Other concepts, like the Doppler effect, require more refined considerations because points
of different trajectories are involved. Luckily enough (\ref{easydopplergen}) is easily exported to 
GR. In such case the particle travels 
through a geodesic with a $4$-velocity computed either wrt proper time for massive 
particles or wrt an affine parameter for massless ones. 
Its velocity $U$ is typically different when evaluated at the Source location, $U_s$,  
than when evaluated at the point of reception by the Observer, $U_o$. Unlike the massive case, the 
affine parameter for photons
is only determined up to an arbitrary constant factor, the consequence being that whereas for 
the massive case both scalar 
products, $V_s\cdot U_s$ and $V_o\cdot U_o$, are invariant\footnote{We mean invariants under 
general changes of coordinates. In the 
passive interpretation of diffeomorphisms a scalar computed 
at a given point becomes an invariant, 
in the sense that its value is independent of the coordinates used to describe such point 
\cite{Pons:2017ljz}.}, it is only their quotient which is invariant for massless particles 
\beq
\frac{E_s}{E_o}= \frac{V_s\cdot U_s}{V_o\cdot U_o} \,.
\label{easydopplergen2}
\eeq
The above expression captures the Doppler effect and its extension to the massive case as a ratio 
between some data 
from the emission event,
$V_s\cdot U_s$, to some data from the reception event, 
$V_o\cdot U_o$\footnote{One can go one step further
and parallel transport the data form the Source to the Observer's location. Since $U_s$ is 
transported to $U_o$, one can see that the whole
effect originates from the fact that $V_s$ is not transported to $V_o$.}. 
{\sl We submit that (\ref{easydopplergen2}) must be taken as the definition of the Doppler 
effect --interpreted as an energy shift and also extended to the massive case-- in GR}.  

Let us notice that in adopting  (\ref{easydopplergen2}), hence including a 
computational prescription, for the evaluation of the GR Doppler effect, 
 the notion of the relative velocity between Source and Observer, which is crucial in the SR
 derivation,  has disappeared.

\section{ Application to massive particles }
\label{massDop}
 
%%%%%%%%%%%%%%%%%%%%%%%%%%%%%%%%%%%%%%%%%%%%%
\subsection{ The cosmological energy shift for massive particles}
%%%%%%%%%%%%%%%%%%%%%%%%%%%%%%%%%%%%%%%%%%%%%

We continue in the cosmological FLRW setting (\ref{gencosm}), but 
considering the emission of a massive particle  from a comoving Source located at $r=0$. 
Its geodesic trajectory and velocity wrt proper time, $s$, are
\beq
X(t)= \Big(t, r(t),\theta_0,\phi_0\Big)\,, \quad
U(t)= h(t)\Big(1, r'(t),0,0\Big)\,,
\label{Ut}
\eeq
where $\displaystyle h(t):= \frac{d\,t}{d\,s}$  is obtained by requiring $U(t)^2=-1$,
\beq
h(t)=  \left(1-\frac{a(t)^2 r'(t)^2}{1-\sigma \, r(t)^2}\right)^{-1/2}\,.
\label{dt-wrt-ds}
\eeq

To compute $r(t)$, we formulate the geodesic equation (\ref{geodeq}) with the normalized velocity 
 (\ref{Ut}), obtaining
\beq
r(t)= \frac{1}{\sqrt{\sigma }}\sin \left(\sqrt{\sigma } \int_0^t \frac{1}{a(\tau)
\sqrt{1+ C\, a(\tau)^2}} \, d\tau\right)\,,
\label{r-of-t-massive}
\eeq
with $C>0$ an integration constant related to the initial condition $r'(0)$. 
Similarly to (\ref{r-of-t}), (\ref{r-of-t-massive}) holds also for null or negative $\sigma$. 
In addition, the massless case can be recovered 
in the limit $C\to 0$.

With this at hand the expression for the proper $4$-velocity\footnote{Notice that 
although expressed in terms of the cosmological time it
is a proper velocity, $U(t)^2=-1$.} becomes 
\beq
U(t)= \frac{1}{m}\Big(\sqrt{m^2+\frac{a(0)^2}{a(t)^2}\, p^2},\
\frac{ a(0)}{a(t)^2}\,   p\, \cos \left(\sqrt{\sigma }\, \int_0^t \frac{1}{a(\tau)\sqrt{1+m^2\frac{a(\tau)^2}{a(0)^2 p^2}}} \, d\tau\right)
,0,0\Big)\,,
\label{U-of-t-massive}
\eeq
where, for convenience, we restored the mass $m$ of the particle and defined $p$ through 
$\d C= \Big(\frac{m}{a(0)\, p}\Big)^2$.

Let's interpret the invariants: 
\begin{itemize}
\item[{\sl i)}] At the particular time of emission $t=0$ the energy of the massive particle is given by
\beq
E_s= V_s\cdot (m\,U(0)) =\sqrt{m^2+ p^2}\,.
\eeq
Thus $p$ is interpreted as the initial momentum of the particle as measured by the Source
\footnote{We elaborate on this interpretation in the next subsection.}. 
\item[{\sl ii)}] At the time $t$ of reception we obtain the energy of the particle from the invariant  
\beq
E_o= V_o\cdot (m\,U(t)) = \sqrt{m^2+
 \frac{a(0)^2}{a(t)^2} p^2} \equiv \sqrt{m^2+p(t)^2}\,,\ {\rm with} \quad p(t):=  \frac{a(0)}{a(t)} p\,, 
 \label{pt}
 \eeq
and $p(t)$ is interpreted as the momentum of the particle a time $t$, as as measured by the Observer.
 \end{itemize}
 
%%%%%%%%%%%%%%%%%%%%%%%%%%%%%%%%%%%%%%%%%%%%%
\subsection{The necessary connection between two scales}
%%%%%%%%%%%%%%%%%%%%%%%%%%%%%%%%%%%%%%%%%%%%%
The result  (\ref{U-of-t-massive}) has been obtained using the cosmological FLRW metric 
(\ref{gencosm}), which works at the cosmological scale. 
Evidently if we just consider a small region in the close neighborhood of the Source there is no 
trace of the homogeneous Hubble flow that sources the background and therefore (\ref{gencosm}) is not 
applicable at this scale.
Instead, what is applicable at this local scale are the kinematic relations of SR, as it is stated 
by the equivalence principle. 
But then the question arises as how can we proceed in order to connect both settings, 
cosmological and local one. It is not a coordinate 
transformation because we are talking about different metrics: on one side, the broad-brush 
FLRW metric, obtained by averaging the density of 
matter-radiation on very large scales and assuming homogeneity; on the other side, the approximate 
SR Minkowski metric that holds in every small
neighborhood of spacetime. Both pictures are correct, the only caveat being, as said, that they 
are applicable at completely different scales. 

To our understanding, there is a unique way to physically connect the two scales: 
{\sl one must retain the values of the invariants found above 
when moving from the cosmological scale description to the local SR one, or vice versa}. 
Now made explicit, this is the assumption that was already implicit in the previous subsection.

In the local SR frame at the Source we have $V_{s} =(1,\vec 0)$ and 
$\d U(0) = \frac{1}{m}\Big(\sqrt{m^2+ p^2},  \vec p\Big)$ with $\vec p= m \frac{\vec v}{\sqrt{1-v^2}}$ so that
$V_s\cdot U(0) = \frac{1}{m}\sqrt{m^2+ p^2}$. Analogously we will have at the point of reception, 
at time $t$,
$V_o\cdot U(t) = \frac{1}{m}\sqrt{m^2+ p(t)^2}$, Thus
\beq
\frac{E_s}{E_o}= \frac{V_s\cdot U(0)}{V_o\cdot U(t)} = \frac{\sqrt{m^2+ p^2}}{\sqrt{m^2+ p(t)^2}} = 
\frac{\sqrt{m^2+ p^2}}{\sqrt{m^2+ p^2\Big(\frac{a(0)}{a(t)} \Big)^2}}\,.
\label{easydopplergen3}
\eeq
Notice that in the massless limit, $m\to 0$, we recover (\ref{1+z massless}), and the same happens for large momentum, $p\to\infty$,  as well.

Once established the connection between the two scales, we infer that 
$p$ is indeed the momentum of the particle as seen by the comoving Source at the time of emission, 
and $p(t)$ is indeed the momentum of the particle as seen by the comoving Observer
at the time of reception. Thus, independently of the particle being massive or massless, 
the following relation always holds
\beq
 p(t)\, a(t)= p(0)\, a(0)\,.
\eeq

%%%%%%%%%%%%%%%%%%%%%%%%%%%%%%%%%%%%%%%%%%%%%%%%%
\subsection{The gravitational energy shift for massive particles}
%%%%%%%%%%%%%%%%%%%%%%%%%%%%%%%%%%%%%%%%%%%%%%%%%

Similarly to the massless case, sending and receiving a massive particle will also exhibit a 
shift in its kinetic energy. With the same setup of subsection \ref{gravredshift}, and 
working directly with proper time $s$, the trajectory and velocity will be denoted as
\beq
X(s)= (T(s), R(s),\theta_0,\phi_0)\,,\quad
U(s)=(T'(s), R'(s),\theta_0,\phi_0)\,,
\eeq 
with $U(s)^2=-1$, which implies  
$\d T'(s) = {f(R(s))}^{-1}\sqrt{f(R(s)) + (R'(s))^2}$. The geodesic equation gives
$
f'(R(s)) + 2 R''(s)=0\,,
$
which can be integrated to $f(R(s)) + (R'(s))^2= G^2$, with $G > \sqrt{f(R(s))}$  a constant
related with the initial conditions. Hence 
\beq
U(s)=(T'(s), R'(s),\theta_0,\phi_0)=\Big( \frac{G}{f(R(s))}, \sqrt{G^2-f(R(s))},0,0\Big)\,.
\label{UwithG}
\eeq
If we set  $s=0$ for $T=0$  and using  (\ref{Vo})
one can compute the invariant 
$$V_{o}(s)\cdot U(s)= \frac{G}{\sqrt{f(R(s))}}\,,
$$
which implies
\beq
\frac{E(R_0)}{E(R_1)}=
\frac{V_{o}(0)\cdot U(0)}{V_{o}(s)\cdot U(s)} = \sqrt{\frac{f(R_1)}{f(R_0)}}\,,
\label{zgrav2}
\eeq
where $E(R_0)$ is the kinetic energy of the emitted particle as seen by the Source and  $E(R_1)$ 
is the kinetic energy of the received particle as seen by 
the Observer. Unlike the cosmological case above, this result is independent of the mass of the 
particle and it directly admits the massless limit yielding (\ref{zgrav}) in which case these energy ratios can also be read as quotients of frequencies of the photon.

\end{document}